\documentclass[12pt]{article}
\usepackage{fullpage}
\usepackage[centertags]{amsmath}
\usepackage{amssymb}
\usepackage{bm}
\usepackage{cite}
\usepackage{slashed}
\usepackage{url}
\usepackage[hyperindex]{hyperref}
\begin{document}

\title{On the implementation of CVC in weak charged-current proton-neutron transitions}

\author{\textbf{C. Giunti}
\\[0.1cm]
INFN, Sezione di Torino, Via P. Giuria 1, I--10125 Torino, Italy
}

\date{\small(30 January 2016)}

\maketitle

\begin{abstract}
It is shown that the standard expression
of the vector part of the hadronic matrix element
in weak charged-current proton-neutron transitions
is in agreement with the CVC hypothesis,
contrary to a different claim in a recent paper.
\end{abstract}

It has been argued in Ref.~\cite{Ankowski:2016oyj} that the conserved vector current (CVC)
hypothesis
\cite{Feynman:1958ty}
implies a vector current in weak charged-current proton-neutron transitions which is different than the usual one
(see, for example, Ref.~\cite{Giunti:2007ry}).

Let us consider the inverse neutron decay process considered in Ref.~\cite{Ankowski:2016oyj}:
\begin{equation}
\bar\nu_{e} + p \to n + e^{+}
.
\label{ind}
\end{equation}
The vector part of the hadronic matrix element
has the general form
(see, for example, Ref.~\cite{Giunti:2007ry})
\begin{equation}
\langle n(p_{n}) |
v^{\mu}(0)
| p(p_{p}) \rangle
=
\overline{u_{n}}(p_{n})
\left[
\gamma^{\mu} \, F_{1}(q^{2})
+
\frac{ i \, \sigma^{\mu\eta} \, q_{\eta} }{ 2 \, m_{N} } \, F_{2}(q^{2})
+
\frac{ q^{\mu} }{ m_{N} } \, F_{3}(q^{2})
\right]
u_{p}(p_{p})
,
\label{m1}
\end{equation}
where
$m_{N} \simeq 939 \, \text{MeV}$ is the average nucleon mass,
$q = p_{n} - p_{n}$
is the four momentum transfer
and $v^{\mu}(x)$
is the vector part of the quark charged current:
\begin{equation}
v^{\mu}(x)
=
\overline{d}(x) \gamma^{\mu} u(x)
.
\label{qcc}
\end{equation}
The form factors
$F_{1}(q^{2})$,
$F_{2}(q^{2})$ and
$F_{3}(q^{2})$
are called,
respectively,
vector,
weak magnetism and
scalar.
The scalar form factor
$F_{3}(q^{2})$
is generated by a second-class current
\cite{Weinberg:1958ut}
and is well-known to vanish under the CVC hypothesis,
which is a consequence of the invariance of strong interactions
under isospin transformations
(see, for example, Ref.~\cite{Giunti:2007ry}).
The scalar form factor
$F_{3}(q^{2})$
has also been severely limited experimentally.
A recent survey of superallowed nuclear $\beta$ decays
in which $q^{2}$ is very small
found
\cite{Hardy:2014qxa}
\begin{equation}
|F_{3}(0)|
<
0.0035
\,
\frac{m_{N}}{m_{e}}
\,
|F_{1}(0)|
\qquad
\text{(90\% C.L.)}
.
\label{f3exp}
\end{equation}

In Ref.~\cite{Ankowski:2016oyj}
second class currents are claimed to be neglected,
but it is argued that the CVC hypothesis
leads to an additional term
multiplying $F_{1}(q^{2})$ in Eq.~(\ref{m1}).
It is shown in the following that this additional term is just the second class
current that was supposed to be neglected,
leading to a contradiction.

The implementation of the CVC hypothesis in Ref.~\cite{Ankowski:2016oyj}
is done by assuming the current conservation relation
\begin{equation}
\partial_{\mu} v^{\mu}(x) = 0
,
\label{cvce1}
\end{equation}
which implies
\begin{equation}
\langle n(p_{n}) |
q_{\mu} v^{\mu}(0)
| p(p_{p}) \rangle
=
0
.
\label{cvce2}
\end{equation}
Then, from Eq.~(\ref{m1}) we get the constraint
\begin{equation}
F_{3}(q^2)
=
-
\frac{m_{N}}{q^2}
\left( m_{n} - m_{p} \right)
F_{1}(q^2)
.
\label{cvce3}
\end{equation}
Inserting this value of $F_{3}$ in Eq.~(\ref{m1}),
we obtain
\begin{equation}
\langle n(p_{n}) |
v^{\mu}(0)
| p(p_{p}) \rangle
=
\overline{u_{n}}(p_{n})
\left[
\left(
\gamma^{\mu}
-
\frac{q^{\mu}}{q^2}
\slashed{q}
\right)
F_{1}(q^{2})
+
\frac{ i \, \sigma^{\mu\eta} \, q_{\eta} }{ 2 \, m_{N} } \, F_{2}(q^{2})
\right]
u_{p}(p_{p})
.
\label{m2}
\end{equation}
This is the expression for the matrix element
$
\langle n(p_{n}) |
v^{\mu}(0)
| p(p_{p}) \rangle
$
which is claimed in Ref.~\cite{Ankowski:2016oyj} to be correct under the CVC hypothesis.

Note the contradiction with the initial
statement in Ref.~\cite{Ankowski:2016oyj} that the contribution $F_{3}$
of second class currents is neglected.
Moreover, it is clear that Eq.~(\ref{m2}) cannot be correct,
because:

\begin{enumerate}

\item
The CVC hypothesis based on the invariance of strong interactions
under isospin transformations implies that
$F_{3}(q^2)=0$,
not the value in Eq.~(\ref{cvce3}).

\item
The value of $F_{3}(0)$ in Eq.~(\ref{cvce3})
diverges at $q^2 \to 0$,
in sharp contradiction with the experimental bound in Eq.~(\ref{f3exp}).
For example,
in neutron decay we have
$q^2 \sim (m_{n} - m_{p})^2$
and
Eq.~(\ref{cvce3}) would give
$|F_{3}(0)| \sim 0.4 m_{N} |F_{1}(0)| / m_{e}$.

\end{enumerate}

Then, one can ask what is wrong with the argument in Ref.~\cite{Ankowski:2016oyj}.
The mistake is in assuming the exact validity of the current conservation relation
(\ref{cvce1}),
which is not correct,
because isospin symmetry is broken by the mass difference of the $u$ and $d$ quarks and
by electromagnetic interactions.
Indeed, one can find that
\begin{equation}
\partial_{\mu} v^{\mu}(x)
=
i \left( m_{d} - m_{u} \right)
\overline{d}(x) u(x)
-
i e \overline{d}(x) \slashed{A}(x) u(x)
,
\label{cvcb}
\end{equation}
where $e$ is the elementary electric charge and $A^{\mu}(x)$
is the electromagnetic field.

Therefore,
one can use the current conservation relation (\ref{cvce1})
only in the approximation of exact isospin invariance,
which is equivalent to neglect the difference of the
proton and neutron masses in Eq.~(\ref{cvce3}),
leading to the correct well-known CVC result $F_{3}(q^2)=0$.

In conclusion,
I have shown that the standard expression
of the vector part of the hadronic matrix element
in weak charged-current proton-neutron transitions
is in agreement with the CVC hypothesis,
contrary to the claim in Ref.~\cite{Ankowski:2016oyj}.

\bibliographystyle{physrev3}
\bibliography{cvc}

\begin{thebibliography}{1}

\bibitem{Ankowski:2016oyj}
A.~M. Ankowski,
arXiv:1601.06169.

\bibitem{Feynman:1958ty}
R.~P. Feynman and M.~Gell-Mann,
Phys. Rev. {\bf 109}, 193 (1958).

\bibitem{Giunti:2007ry}
C.~Giunti and C.~W. Kim,
{\em {Fundamentals of Neutrino Physics and Astrophysics}} (Oxford
University Press, Oxford, UK, 2007).

\bibitem{Weinberg:1958ut}
S.~Weinberg,
Phys. Rev. {\bf 112}, 1375 (1958).

\bibitem{Hardy:2014qxa}
J.~C. Hardy and I.~S. Towner,
Phys. Rev. {\bf C91}, 025501 (2015), arXiv:1411.5987.

\end{thebibliography}

\end{document}